\documentstyle[prd,aps,12pt,epsf,psfig]{revtex}
\begin{document}

\baselineskip=7mm
\def\ap#1#2#3{           {\it Ann. Phys. (NY) }{\bf #1} (19#2) #3}
\def\arnps#1#2#3{        {\it Ann. Rev. Nucl. Part. Sci. }{\bf #1} (19#2) #3}
\def\cnpp#1#2#3{        {\it Comm. Nucl. Part. Phys. }{\bf #1} (19#2) #3}
\def\apj#1#2#3{          {\it Astrophys. J. }{\bf #1} (19#2) #3}
\def\asr#1#2#3{          {\it Astrophys. Space Rev. }{\bf #1} (19#2) #3}
\def\ass#1#2#3{          {\it Astrophys. Space Sci. }{\bf #1} (19#2) #3}

\def\apjl#1#2#3{         {\it Astrophys. J. Lett. }{\bf #1} (19#2) #3}
\def\ass#1#2#3{          {\it Astrophys. Space Sci. }{\bf #1} (19#2) #3}
\def\jel#1#2#3{         {\it Journal Europhys. Lett. }{\bf #1} (19#2) #3}

\def\ib#1#2#3{           {\it ibid. }{\bf #1} (19#2) #3}
\def\nat#1#2#3{          {\it Nature }{\bf #1} (19#2) #3}
\def\nps#1#2#3{          {\it Nucl. Phys. B (Proc. Suppl.) } {\bf #1} (19#2) #3}
\def\np#1#2#3{           {\it Nucl. Phys. }{\bf #1} (19#2) #3}

\def\pl#1#2#3{           {\it Phys. Lett. }{\bf #1} (19#2) #3}
\def\pr#1#2#3{           {\it Phys. Rev. }{\bf #1} (19#2) #3}
\def\prep#1#2#3{         {\it Phys. Rep. }{\bf #1} (19#2) #3}
\def\prl#1#2#3{          {\it Phys. Rev. Lett. }{\bf #1} (19#2) #3}
\def\pw#1#2#3{          {\it Particle World }{\bf #1} (19#2) #3}
\def\ptp#1#2#3{          {\it Prog. Theor. Phys. }{\bf #1} (19#2) #3}
\def\jppnp#1#2#3{         {\it J. Prog. Part. Nucl. Phys. }{\bf #1} (19#2) #3}

\def\rpp#1#2#3{         {\it Rep. on Prog. in Phys. }{\bf #1} (19#2) #3}
\def\ptps#1#2#3{         {\it Prog. Theor. Phys. Suppl. }{\bf #1} (19#2) #3}
\def\rmp#1#2#3{          {\it Rev. Mod. Phys. }{\bf #1} (19#2) #3}
\def\zp#1#2#3{           {\it Zeit. fur Physik }{\bf #1} (19#2) #3}
\def\fp#1#2#3{           {\it Fortschr. Phys. }{\bf #1} (19#2) #3}
\def\Zp#1#2#3{           {\it Z. Physik }{\bf #1} (19#2) #3}
\def\Sci#1#2#3{          {\it Science }{\bf #1} (19#2) #3}

\def\n.c.#1#2#3{         {\it Nuovo Cim. }{\bf #1} (19#2) #3}
\def\r.n.c.#1#2#3{       {\it Riv. del Nuovo Cim. }{\bf #1} (19#2) #3}
\def\sjnp#1#2#3{         {\it Sov. J. Nucl. Phys. }{\bf #1} (19#2) #3}
\def\yf#1#2#3{           {\it Yad. Fiz. }{\bf #1} (19#2) #3}
\def\zetf#1#2#3{         {\it Z. Eksp. Teor. Fiz. }{\bf #1} (19#2) #3}
\def\zetfpr#1#2#3{         {\it Z. Eksp. Teor. Fiz. Pisma. Red. }{\bf #1} (19#2) #3}
\def\jetp#1#2#3{         {\it JETP }{\bf #1} (19#2) #3}
\def\mpl#1#2#3{          {\it Mod. Phys. Lett. }{\bf #1} (19#2) #3}
\def\ufn#1#2#3{          {\it Usp. Fiz. Naut. }{\bf #1} (19#2) #3}
\def\sp#1#2#3{           {\it Sov. Phys.-Usp.}{\bf #1} (19#2) #3}
\def\ppnp#1#2#3{           {\it Prog. Part. Nucl. Phys. }{\bf #1} (19#2) #3}
\def\cnpp#1#2#3{           {\it Comm. Nucl. Part. Phys. }{\bf #1} (19#2) #3}
\def\ijmp#1#2#3{           {\it Int. J. Mod. Phys. }{\bf #1} (19#2) #3}
\def\ic#1#2#3{           {\it Investigaci\'on y Ciencia }{\bf #1} (19#2) #3}
\def\tp{these proceedings}
\def\pc{private communication}
\def\ip{in preparation}
\newcommand{\TeV}{\,{\rm TeV}}
\newcommand{\GeV}{\,{\rm GeV}}
\newcommand{\MeV}{\,{\rm MeV}}
\newcommand{\keV}{\,{\rm keV}}
\newcommand{\eV}{\,{\rm eV}}
\newcommand{\Tr}{{\rm Tr}\!}
\renewcommand{\arraystretch}{1.2}
\newcommand{\be}{\begin{equation}}
\newcommand{\ee}{\end{equation}}
\newcommand{\bea}{\begin{eqnarray}}
\newcommand{\eea}{\end{eqnarray}}
\newcommand{\ba}{\begin{array}}
\newcommand{\ea}{\end{array}}
\newcommand{\bmat}{\left(\ba}
\newcommand{\emat}{\ea\right)}
\newcommand{\refs}[1]{(\ref{#1})}
\newcommand{\ler}{\stackrel{\scriptstyle <}{\scriptstyle\sim}}
\newcommand{\ger}{\stackrel{\scriptstyle >}{\scriptstyle\sim}}
\newcommand{\lag}{\langle}
\newcommand{\rag}{\rangle}
\newcommand{\ns}{\normalsize}
\newcommand{\cm}{{\cal M}}
\newcommand{\gr}{m_{3/2}}
\newcommand{\p}{\partial}
\renewcommand{\le}{\left(}
\newcommand{\ri}{\right)}
\relax
\def\321{$SU(3)\times SU(2)\times U(1)$}
\def\21{$SU(2)\times U(1)$}
\def\dbd{0\nu\beta\beta}
\def\ord{{\cal O}}
\def\tl{{\tilde{l}}}
\def\tL{{\tilde{L}}}
\def\bd{{\overline{d}}}
\def\tL{{\tilde{L}}}
\def\a{\alpha}
\def\b{\beta}
\def\g{\gamma}
\def\c{\chi}
\def\d{\delta}
\def\D{\Delta}
\def\db{{\overline{\delta}}}
\def\Db{{\overline{\Delta}}}
\def\e{\epsilon}
\def\l{\lambda}
\def\n{\nu}
\def\m{\mu}
\def\nt{{\tilde{\nu}}}
\def\p{\phi}
\def\P{\Phi}
\def\solm{\Delta_{\odot}}
\def\sola{\theta_{\odot}}
\def\mee{m_{ee}}
\def\atm{\Delta_{\makebox{\tiny{\bf atm}}}}
\def\k{\kappa}
\def\x{\xi}
\def\r{\rho}
\def\s{\sigma}
\def\t{\tau}
\def\th{\theta}
\def\om{\omega}
\def\ne{\nu_e}
\def\nm{\nu_{\mu}}
\def\snui{\tilde{\nu_i}}
\def\ehat{\hat{e}}
\def\la{{\makebox{\tiny{\bf loop}}}}
\def\ta{\tilde{a}}
\def\tb{\tilde{b}}
\def\mb{m_{1b}}
\def\mt{m_{1 \tau}}
\def\rl{{\rho}_l}
\def\meg{\m \rightarrow e \g}

\renewcommand{\Huge}{\Large}
\renewcommand{\LARGE}{\Large}
\renewcommand{\Large}{\large}
\title{ Radiative origin of solar scale and $U_{e3}$}
\author{ Anjan S. Joshipura \\[.5cm]
{\ns\it  Theoretical Physics Group, Physical Research Laboratory,}\\
{\ns\it Navarangpura, Ahmedabad, 380 009, India.}} \maketitle
\vskip .5cm
\begin{center}
{\bf Abstract}
\end{center}
\vskip.5cm We make a general study of possibility of  generating
solar scale $\solm$ and the CHOOZ angle $U_{e3}$ radiatively by
assuming that they are zero at some high scale. The most general
neutrino mass matrix leading to this result is determined in a CP
conserving theory. This matrix contains four independent
parameters which can be fixed in terms of physical observables.
The standard weak radiative corrections then lead to non-zero
$\solm$ and $U_{e3}$ without drastically altering the other tree
level results. As a consequence, both $\solm$ and $U_{e3}$ are
predicted in terms of other physically observable parameters.
These predictions  are insensitive to specific form of the
neutrino mass matrix. The solar scale and $U_{e3}$ are strongly
correlated with the effective neutrino mass $m_{ee}$ probed in
neutrinoless double beta decay. In particular, the LMA solution to
the solar neutrino problem arise for $m_{ee}$ close to the present
experimental limit. An example of specific texture is presented
which predicts maximal atmospheric mixing and $\tan^2
\theta_{\odot}\approx 0.5$ for the solar mixing angle
$\theta_{\odot}$.
\newpage

{\bf Introduction:} Neutrino oscillation experiments have provided
significant information on neutrino masses and mixing \cite{rev}.
The "standard" picture emerging from analysis of various
experiments is the existence of two hierarchical (mass)$^2$
differences and two large and one small or zero mixing angle among
three neutrinos. The overall scale of neutrino masses is not fixed
directly by neutrino oscillation experiments. This complimentary
information can be obtained from direct neutrino mass measurements
\cite{mainz} and from neutrinoless double beta decay ($\dbd$)
experiments \cite{dbd}. These types of experiments have so far
provided only upper limits. The neutrino mass (assuming no mixing)
is constrained by tritium $\b$ decay experiments to be less than
2.2 eV. The best limit from $\dbd$ decay experiments is
\be \label{mee1}
|m_{ee}|\equiv |\sum U_{ei}^2 m_{\nu_i}|\leq 0.38 ~h
\eV~~~~~~~~{\rm at~95\%~ CL}~,\ee
where $h\sim 0.6-2.8$ denotes the uncertainty in nuclear matrix
element \cite{vissani}. $U$ denotes here the neutrino mixing
matrix and $m_{\nu_i} (i=1,2,3)$ are neutrino mass values which
can take either sign.

While the hierarchical neutrino masses cannot be ruled out at
present, the presence of large mixing angles hints at an almost
degenerate pair of neutrinos. This will become a necessity if
$m_{ee}$ would be found to be significantly larger than the
atmospheric scale. This would require all three neutrinos to be
nearly degenerate \cite{deg} if mixing among them is also to
account for the solar and atmospheric neutrino results. The  mass
patterns with two \cite{2deg} or all the three
\cite{deg,rad1,rad2} nearly degenerate neutrinos are therefore of
considerable importance.

If overall scale of neutrino masses is larger than the atmospheric
scale then one would like to understand why neutrino (mass)$^2$
differences (particularly, the solar scale $\solm$)  are much
smaller? Interesting possibility is to assume degenerate neutrinos
which get split \cite{deg,rad1,rad2} by radiative corrections
\cite{rg,antu} induced through charged current interactions in the
standard model (SM) or in the minimal supersymmetric standard
model (MSSM). Advantage of this scheme is its predictive power.
The most general $3\times 3$ neutrino mass matrix with completely
degenerate spectrum is characterized \cite{branco} in terms of two
angles and one phase. These three parameters determine all
neutrino masses and (complex) mixing after the known radiative
corrections are included. Unfortunately, this predictive scenario
does not give \cite{rad2} phenomenologically required description
of neutrino masses and mixing.

The next best possibility is to assume that only two of the three
neutrinos are exactly degenerate at high scale. The radiative
corrections then lead to the solar splitting within this scheme.
It is possible to do a quite general and fairly model independent
analysis of this situation which we present in this paper. We
assume that neutrino masses, the atmospheric scale and two large
mixing angles are tree level effects and are already described by
neutrino mass matrix specified at a high scale. We require this
mass matrix to have vanishing solar scale and vanishing CHOOZ
\cite{chooz} angle in flavour basis. Neutrino mass matrix with
such property can be characterized by four independent parameters
in a CP conserving theory. Solar scale and the CHOOZ angle
$U_{e3}$ arise after radiative corrections and represent generic
prediction of this scheme. These predictions are found to be model
independent and hold for all the matrices under consideration. We
now present this analysis and discuss its consequences.

\noindent{\bf General analysis of pseudo Dirac neutrinos: } Let us
consider a CP conserving theory specified by a general $3\times 3$
real symmetric neutrino mass matrix $M_{\nu 0}$. This matrix can
always be specified in flavor basis corresponding to diagonal
charged lepton masses. We adopt the following general
parameterization for $M_{\nu 0}$ in the flavor basis.
\be \label{para}M_{\nu 0}= \bmat{ccc} s_1&t&u\\
t&s_2&v\\u&v&s_3\\ \emat \ee
We assume that the above $M_{\nu 0}$ describes physics at a high
scale $M_X$. $M_{\nu 0}$ is required to yield vanishing solar
scale and CHOOZ angle at $M_X$. Let us derive conditions on
elements of $M_{\nu 0}$ for this to happen.

Vanishing of the solar scale requires that two of the eigenvalues
of $M_{\nu 0}$ are  degenerate with masses $(m,m)$ or $(m,-m)$
($m>0$). The relative angle between the degenerate pair can be
rotated away in the former case in a CP conserving theory. This is
not true in case of masses differing in their sign. Thus barring
possibility of radiative amplification \cite{radamp}, the former
case will not lead to large solar angle and we concentrate on the
second possibility with masses $(m,-m)$. Such a pair is equivalent
to a Dirac neutrino invariant under some global $U(1)$ symmetry.
The standard weak current would violate this symmetry in general
\cite{wolf} and the Dirac state would split into a pair of pseudo
Dirac neutrinos. General conditions under which this happens in
case of three generations were discussed in \cite{pdp}. In
particular, the $M_{\nu 0}$ should satisfy
\be \label{cond1}
 tr(M_{\nu 0})\sum_i \Delta_i=det M_{\nu 0} ~,\ee where $\Delta_i$
represents the determinant of the $2\times 2$ block of $M_{\nu 0}$
obtained by blocking the $i^th$ row and column.

When the above condition is satisfied, eigenvalues of $M_{\nu 0}$
are given by $(m,-m,~T)$  where
$$ m\equiv \sqrt{-\sum_i{\Delta_i}}~~~~~~~~~~~ T\equiv tr(M_{\nu 0}) $$
Let $U_0$ diagonalize $M_{\nu 0}$:
\be \label{dia}
U_0^T~M_{\nu 0}~U_0=Diag.(m,-m,~T)~\ee

Since $M_{\nu 0}$ is specified in the flavor basis, $U_0$ defined
above represents physical neutrino mixing matrix at tree level.
The electron neutrino survival probability in reactor experiments
such as CHOOZ is given by $(U_0)_{e3}$ which we require to be
zero. One can show that $M_{\nu 0}$ satisfies eq.(\ref{cond1}) and
leads to $(U_0)_{e3}=0$ provided

\bea \label{cond} v^2&=& (s_1+s_2)(s_1+s_3)~; \nonumber \\
t&=&-{u v \over s_1+s_2}~. \eea
The second equation does not hold in a special case with $v=0$. In
this case $t$ and $u$ are unrelated and the above two conditions
uniquely lead to the following $M_{\nu 0}$:
\be \label{ansatz}M_{\nu 0}= \bmat{ccc} s&t&u\\
t&-s&0\\u&0&-s\\ \emat \ee
The detailed phenomenological consequences of the special solution
given in eq.(\ref{ansatz}) were worked out in \cite{pred}. Here we
consider rest of the the neutrino mass matrices specified by
restriction given in eq.(\ref{cond}). Let us parameterize the
mixing matrix $U_0$ by
\be \label{u0} U_0=\bmat{ccc}
c_{\phi_0}&-s_{\phi_0}&0\\ s_{\phi_0} c_{\theta_0}
&c_{\phi_0} c_{\theta_0}&-s_{\theta_0}\\
s_{\phi_0} s_{\theta_0} &s_{\theta_0} c_{\phi_0}&c_{\theta_0}\\
\emat ~,\ee
The mixing angles are determined using eq.(\ref{dia}) and
eq.(\ref{cond}):
\bea\label{ma0} \tan{\phi_0}&=&\sqrt{{m-s_1\over m+s_1}}~,
\nonumber
\\ \tan\theta_0&=&{u\over t}~. \eea
Here, $m$ denotes the common mass of the Dirac pair and is given
by
$$m=\sqrt{s_1^2+t^2+u^2}$$
We note that
\begin{itemize}
\item The effective neutrino mass probed in the $\dbd$
 is given by $$\mee^0=s_1$$
\item At the tree level, there is only one $(mass)^2$ difference
which provides the atmospheric scale \be
\label{atm0}\Delta_{0A}\equiv|m^2-T^2| ~. \ee
Corresponding mixing angle ($\equiv \theta_{A}^0$) coincides with
$\theta_0$ and is large when $t\sim u$:
\be \label{atang0} \sin^22 \theta_A^0=\sin^2 2 \theta_0 ~.\ee
\item There is no solar splitting at this stage but would be solar
mixing angle  $\sola^0$ coincides with  $\phi_0$ and is given by
\be \label{solang0} \tan^2\sola^0={m-\mee^0 \over m+\mee^0} \ee
\end{itemize}

The above relations are valid for the most general $M_{\nu 0}$
with parameters satisfying eq.(\ref{cond}).  They are derived at
tree level but as we demonstrate latter, radiative corrections do
not substantially change them. The major effect of radiative
corrections is to generate the solar splitting and a non-zero
value for $U_{e3}$. It turns out that these quantities are not
arbitrary but are predicted in terms of other observables
irrespective of the detailed form of $M_{\nu 0}$. This happens
because two conditions in eq.(\ref{cond}) leave us with four
independent parameters. They can be determined in terms of four
observables namely, atmospheric scale and angle, effective mass
probed in $\dbd$ and the solar angle. The solar splitting and
$U_{e3}$ generated radiatively then no longer remain arbitrary but
are  determined in terms of these observables.

We can express all parameters of $M_{\nu 0}$ in terms of
observables using conditions of eq.(\ref{cond}):
\be\label{observ}
\ba{cc} m=|{\mee^0\over \cos 2 \sola^0}|&t^2=\cos^2 \theta_A^0~
(m^2-\mee^{0 2})\\
T^2= (m^2- \Delta_{A},m^2+\Delta_{A})&({\mbox for
~m^2>\Delta_{A},<\Delta_A})\\
s_3= \cos^2\theta_A^0 ~T-\sin^2\theta_A^0 ~\mee^0&
s_2=\sin^2\theta_A^0~ T-\cos^2\theta_A^0~ \mee^0
\\ \ea \ee

The solar splitting can be  obtained \cite{pdp} using the relevant
renormalization group equations\cite{rg,antu}. The consequences
of these equations have been discussed in a number of
papers\cite{rad1,rad2}.

The radiatively corrected neutrino mass matrix is given by
\be \label{radmnu} M_{\nu}=I_gI_t\bmat{ccc}I_e^{\frac{1}{2}}&0&0
\\ 0&I_\m^{\frac{1}{2}}&0 \\ 0&0&I_\tau^{\frac{1}{2}}\\\emat~M_{\nu 0}
~\bmat{ccc}I_e^{\frac{1}{2}}&0&0 \\ 0&I_\m^{\frac{1}{2}}&0 \\
0&0&I_\tau^{\frac{1}{2}}\\\emat ~,\ee where

$$I_\a^{\frac{1}{2}}\equiv 1+\delta_a~,$$
with \be \label{deltai}
\delta_\a\approx c({m_\alpha\over 4 \pi v })^2 ln{M_X\over M_Z}~.
\ee $M_X$ here corresponds to a large scale and we take $M_X\sim
10^{16} \GeV$; $c=\frac{3}{2},-\frac{1}{\cos^2\b}$ in respective
cases of the standard model \footnote {The value of $c$ in case of
the standard model is given by 1/2(3/2) according to calculations
in \cite{rg} (\cite{antu}). We will use the value as in
\cite{antu}.} and the minimal supersymmetric standard model
\cite{rad1} and $\a=e,\mu,\tau$. $I_{g,t}$ are calculable
coefficient summarizing the effect of the gauge and the top quark
corrections.

Apart from the overall factor $I_gI_t$, the radiative corrections
are largely dominated by the $\tau$ Yukawa couplings and it is
easy to determine neutrino mixing angle and masses keeping only
$\delta_\tau$ corrections and working to the lowest order in
$\delta_\tau$. We now have
$$U^T~M_\nu~U=Diag.(m_{\nu_1},m_{\nu_2},m_{\nu_3})~,$$
with \bea \label{masses}
m_{\nu_1}&\approx& I_g I_t (m+\delta_\tau \sin^2\theta_A^0 (m-\mee^0))+\ord(\delta_{\tau}^2)~,\nonumber \\
m_{\nu_2}&\approx& I_g I_t (-m-\delta_\tau \sin^2\theta_A^0 (m+\mee^0) )+\ord(\delta_{\tau}^2)~,\nonumber \\
m_{\nu_3}&\approx& I_g I_t (T+2 \delta_\tau T \cos^2\theta_A^0
)+\ord(\delta_{\tau}^2)~, \eea
where we have used eq.(\ref{observ}). The tree level mixing matrix
$U_0$ gets modified to a general $U$:
\be \label{u} U=\bmat{ccc}
c_\phi c_\omega&-s_\p c_\omega&s_\omega\\
c_\p s_\theta s_\omega+c_\theta s_\phi&c_\theta c_\phi-s_\phi s_\theta s_\omega&-s_\theta c_\omega\\
-c_\phi c_\theta s_\omega+s_\phi s_\theta&s_\p s_\omega c_\theta+s_\theta c_\p&c_\theta c_\omega\\
\emat ~,\ee
As before, the angles $\phi,\theta$ correspond respectively to
solar and atmospheric mixing angles. These are now given by
\bea\label{theta} \tan\theta_A&=&\tan\theta_A^0~
\left(1+{\delta_\tau \over m^2-T^2} (m^2+T^2-2 T
s_1)\right)+\ord(\delta_{\tau}^2),
\nonumber \\
\tan^2\sola&=&\tan^2\sola^0+\ord(\delta_{\tau}^2), \eea
where $\theta_A^0$ (eq.(\ref{atang0})) and
$\sola^0$(eq.(\ref{solang0})) are tree level atmospheric and the
solar mixing angles respectively. Note that the solar mixing angle
does not receive radiative corrections to the lowest order in
$\delta_\tau$.

The effective neutrino mass probed in $\dbd$ is now given by  \be
\label{meff} m_{ee}=I_gI_t \mee^0=I_gI_t ~s_1 .\ee
 The atmospheric scale also receive
radiative corrections and is now given by \be
\label{atm}\Delta_{A}\equiv
\frac{1}{2}(m_{\nu_1}^2+m_{\nu_2}^2)-m_{\nu_3}^2=I_g^2I_t^2\left(
\Delta_{0A}+2 \delta_\tau (m^2\sin^2\theta-2 T^2
\cos^2\theta)\right)+\ord(\delta_{\tau}^2) ~. \ee
It is seen that all the tree level predictions receive small
radiative corrections. Thus all the neutrino mass matrices leading
to two degenerate states and characterized by eq.(\ref{cond}) are
stable against radiative corrections. This is to be contrasted
with the case of fully degenerate neutrino spectrum which leads to
radiative instability in some specific cases \cite{rad1}. The
non-trivial effect of the radiative corrections is generation of
the solar splitting and a non-zero $U_{e3}$:
\bea \label{split1}
\solm&\equiv&m_{\nu_2}^2-m_{\nu_1}^2 \approx 4 \mee \delta_\tau
|{\mee\over \cos 2 \sola}| \sin^2\theta_A ~,\nonumber \\
|U_{e3}|&=& |s_\omega|\sim |{\delta_{\tau} T \sin 2 \theta_A
\sqrt{m^2-m_{ee}^2}\over \Delta_A}|~.\eea

The above equations relate the solar scale and CHOOZ angle to
other experimentally determined quantities as can be seen using
eq.(\ref{observ}). Eq.(\ref{split1}) therefore represent basic
prediction of the scheme defined by eq.(\ref{cond}). It is
remarkable that all these matrices lead to unique predictions in
eq.(\ref{split1}) which are insensitive to  specific texture of
the neutrino mass matrix.

Let us now explore consequences of eq.(\ref{split1}). The
atmospheric mixing angle and scale are experimentally
well-determined: $\Delta_A\approx (1.5-5)\cdot 10^{-3}\eV^2$ and
$\sin^2 2 \theta_A\sim 0.8-1$. The solar scale $\solm$ and mixing
$\sola$ are also highly constrained \cite{solb}, particularly
after \cite{sola} the recent neutral current results from SNO
\cite{sno}. Based on the global analysis of all the solar data,
the only solutions allowed at 3$\sigma$ level are the large mixing
angle solution (LMA) and the LOW solution with $\solm\sim
10^{-7}\eV^2$. The allowed ranges of parameters in these cases at
3$\sigma$ are given approximately by by $\solm\approx 3\cdot
10^{-4}-2 \cdot 10^{-5} \eV^2$ and $\tan^2\sola\sim 0.2-0.9$ in
case of the LMA solution and $\solm\approx 3 \cdot 10^{-8}- 1
\cdot 10^{-7} \eV^2$ and $\tan^2\sola\sim 0.4-0.9$ in case of the
LOW solution. Both the small mixing angle and vacuum solutions are
excluded at 3$\sigma$. In particular, the solar mixing angle is
found to be less than $45^0$ in all the preferred solution a fact
which plays an important role in the following.

 The
predicted value of $\solm$ and $U_{e3}$  are different in SM and
MSSM due to different values of  $\delta_\tau$ in these two cases.
In case of SM, $\delta_\tau\sim 10^{-5}$ while it can become
larger for MSSM due to presence of $\tan\b$. More importantly,
sign of $\delta_{\tau}$ is different in these two cases. The
negative values of $\delta_{\tau}$ in case of MSSM makes it
unsuitable for the description of the solar data as we now argue.

In analyzing the solar data, $\solm$ is chosen positive by
convention and mixing angle $\sola$ is allowed to be greater than
45$^0$. The sign of $\solm$ as defined by eq.(\ref{split1}) is
governed by the sign of $\mee$ and $\delta_{\tau}$. The sign of
$\mee$ also determines magnitude of the solar angle through
eq.(\ref{solang0}). Positive (negative)  values of $\mee$ gives a
$\sola$ less (greater) than one. Since $\delta_{\tau}$ is negative
in case of the MSSM one needs  negative $\mee$ to obtain positive
$\solm$ with the result that  $\tan^2 \sola$ becomes greater than
one \footnote{For positive $\mee$ one needs to reverse the role of
$\nu_1$ and $\nu_2$. The relevant $\tan^2\sola$ is inverse of
eq.(\ref{solang0}) and is also greater than 1.}. Since the solar
neutrino results do not allow $\tan^2\sola>1$, MSSM radiative
corrections as a mechanism to generate the solar splitting is
disfavoured in the present context. In contrast, the SM radiative
corrections give $\tan^2\sola<1$ as required for these solutions.
We discuss this case now.

The numerical value of $\solm$ and $U_{e3}$ are correlated both
with $\mee$ as well as with the solar mixing angle. We show this
correlation in Fig.(1) which displays variation in $10^5
{\solm\over \eV^2}$ (solid) and $10^2 U_{e3}$ (dotted) with $\mee$
for typical values of $\tan^2\sola$ needed for the LMA and LOW
solutions. It is seen from the figure that LMA solution can be
obtained for relatively larger value of $\mee$ typically $\mee\geq
0.1 \eV$.  The minimum required value of $\mee$ increases with
decrease in $\tan^2\sola$. The LOW solution require much smaller
but experimentally accessible values \cite{dbd2} of $\mee$ around
$0.05 \eV$.

The predicted values of $U_{e3}$ are generally smaller than the
present limits as well as possible detection \cite{choozm} limit
around $\sim 0.05$ for most ranges in the parameters. But if
$\tan^2\sola$ is $\sim 0.5-0.8$ then $U_{e3}$ is predicted to be
in the range $0.01-0.1$ and is correlated with the LMA solution.

An interesting consequence \cite{bb} of the correlation between
LMA solution and large $m_{ee}$ is as follows. For
$\tan^2\sola\sim 0.2-0.7$ and $\mee\sim 0.3 \eV$, the common mass
$m=|{\mee\over \cos 2\sola}|$ of the degenerate pair lies in the
range
$$m\approx  0.5-1.8 \eV ~.$$
The third mass $T$ is also required to be close to $m$ since
$\Delta_A\sim |m^2-T^2|$. As a consequence, the LMA solution in
these models is automatically correlated with almost degenerate
neutrino mass spectrum with a common mass much larger than the
atmospheric scale.

The above discussion is based on general class of matrices leading
to pseudo-Dirac neutrino. We now supplement this with a discussion
of a specific texture.

\noindent{\bf Almost degenerate neutrinos:} Consider the following
specific texture:
\be \label{texture}M_{\nu 0}=s \bmat{ccc}1+\e &-2&2\\
-2&1-\e&2\\2&2&1-\e\\ \emat \ee
The above texture is determined by only two parameters $s$ and
$\e$. It satisfies conditions in eq.(\ref{cond}). Thus it leads to
two degenerate eigenvalues and vanishing $U_{e3}$ at a high scale.
The eigenvalues of the above matrix are given by $(m,-m,s(3-\e))$
with
\be \label{mtext} m=s\sqrt{9+2 \e+\e^2}\ee
It is seen that all neutrinos are degenerate when $\e=0$. It is
known \cite{rad2} that matrix with degenerate neutrinos cannot
lead to the required mass pattern after radiative corrections.
This is remedied here by introduction of a small $\e$ which leads
to the atmospheric neutrino splitting at high scale:
\be \label{attext}\Delta_A\approx \Delta_{A0}=8 \e s^2 \ee

The above specific texture has four predictions. As in the general
case, the radiatively generated solar scale and $U_{e3}$ are
predictions of the model. In addition, both the solar and
atmospheric mixing angles instead of being arbitrary are fixed
here by the specific texture.

The solar splitting follows from the general expression in
eq.(\ref{split1}):
\be \label{dstext}\solm\sim 2 \delta_\tau s^2 (1+\e) \sqrt{9+2
\e+\e^2} \ee Two parameters $s$ and $\e$ get determined by the
values of $\solm$ and $\Delta_A$. In particular, relatively large
$\solm\sim 10^{-5}\eV^2$ needs small $\e$ and large $s$. The solar
mixing angle is obtained using eqs.(\ref{solang0},\ref{mtext}) and
is given  in the small $\e$ limit by
\be \tan^2\sola\approx 0.5+O(\e) \ee

Thus this texture automatically predicts large mixing angle which
is in the range required for the LMA or LOW solutions. The
atmospheric mixing angle is predicted to be maximal.
Eq.(\ref{texture}) therefore provides an example of the bi-large
mixing patterns with almost degenerate neutrinos.

One can determine required value of $s,\e$ from
eqs.(\ref{attext},\ref{dstext}). Choosing $\solm\sim 5 \cdot
10^{-5}\eV^2$ and $\Delta_A=3\cdot 10^{-3}$, one finds
$$ \e\sim 1.8 \cdot 10^{-3}~~;~~ s=0.45 \eV~.$$
Thus one can obtain $\solm$ around $10^{-5} \eV^2$ provided the
effective neutrino mass $\mee=s\sim 0.4 \eV$. All the three
neutrinos are almost degenerate with a common mass $1.3 \eV$ which
 is not very far from the present experimental limit
\cite{mainz}.

\noindent{\bf Summary:} We discussed possibility of explaining
small values of the solar scale and the angle $U_{e3}$ through
radiative corrections by assuming that these are zero at a high
scale. Restrictions to be satisfied by neutrino mass matrix for
this purpose in the flavor basis were determined (eq.\ref{cond}).
Since neutrino mass matrix can always be expressed in the flavor
basis, eq.(\ref{cond}) provides general conditions for vanishing
of $U_{e3}$ and solar scale in any model.

We showed that the standard weak radiative corrections lead to the
solar splitting required on phenomenological grounds. Both $\solm$
and $U_{e3}$ are predicted in terms of other observables. These
predictions are remarkably independent of detailed form of $M_{\nu
0}$ and remain true for any $M_{\nu 0}$ satisfying
eq.(\ref{cond}).

Detailed analysis presented here shows that one can obtain the
most preferred LMA solution for $\mee\gtrsim 0.1 \eV$. Thus
verification of LMA solution and moderate improvement in limit on
$\mee$ can rule out the entire class of solutions proposed here.
The LMA solution also gets correlated in these models with almost
degenerate neutrino mass spectrum and measurably large $U_{e3}\sim
.01-0.1$.

It is not possible to obtain the correct solar parameters if
radiative corrections are induced in MSSM. It may be possible to
make MSSM also viable by allowing  some nonzero $U_{e3}$ at high
scale and/or by invoking additional sources of radiative
corrections \cite{valle}.\\[.5cm]
{\bf Acknowledgments:} I thank Saurabh Rindani for his interest
in this work and for helpful discussions.\\

\newpage
\begin{figure}[h]
\centerline{\psfig{figure=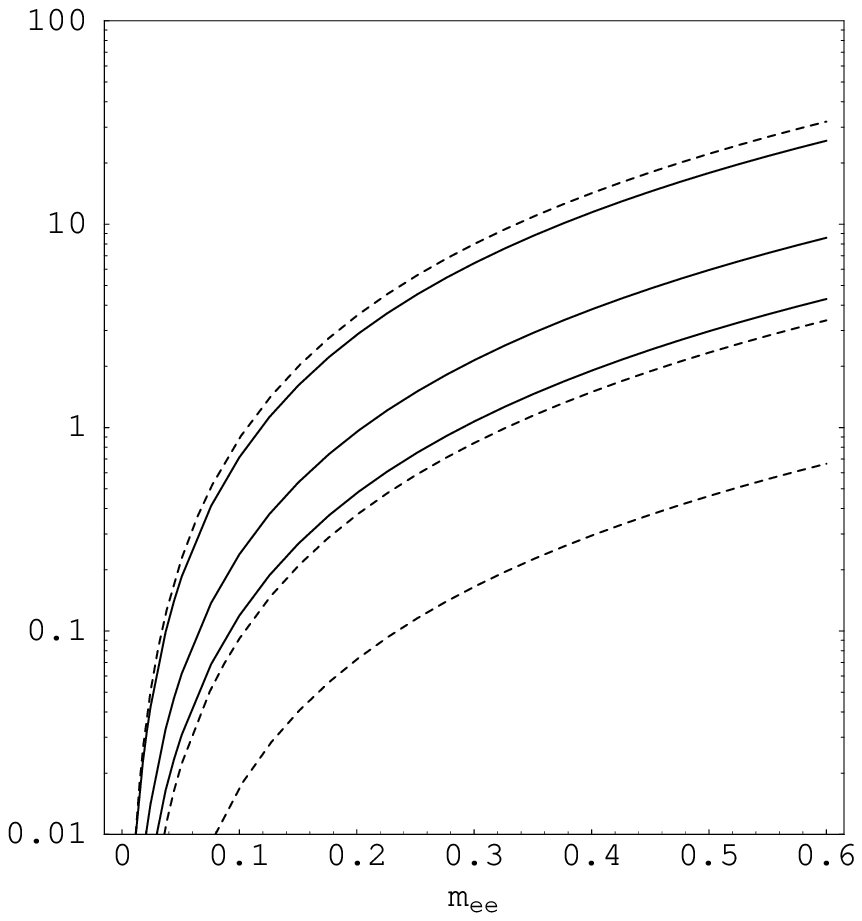,height=15cm,width=15cm}}
\vskip.25cm \caption{ $10^5 \solm$ in ($\eV^2$) (solid) and
$10^2|U_{e3}|$ (dotted) shown as a function of $\mee$ (in $\eV$)
for various values of $tan^2\sola$. The upper middle and lower
curves for each quantities correspond to $\tan^2\sola=0.8,0.5$ and
$0.2$ respectively. }
\end{figure}

\end{document}